# EMERGING TECHNOLOGIES AND METHODS IN WIDE-AREA SEARCH FOR NUCLEAR MATERIALS

L.E. SINCLAIR
Natural Resources Canada
Ottawa, Canada
Email: laurel.sinclair@nrcan-rncan.gc.ca

D.A. MCCORMACK
Natural Resources Canada
Ottawa, Canada

**Abstract**

Canada, a Tier 1 nuclear nation involved in uranium mining and refining, operating nuclear power reactors, and with a Small Modular Reactor action plan, maintains a rigorous nuclear security infrastructure. The Nuclear Emergency Response team at Natural Resources Canada fulfills federal mandates in high-sensitivity air- and ground-based mobile survey for prevention, detection and response. A robust operational framework exists for deployment of traditional large-volume NaI(Tl)-based detection suites. At the same time, a research arm examines emerging non-nuclear technologies which can enhance the capabilities of the operational team. Herein, the potential for uncrewed mobile systems in nuclear security and emergency response operations is discussed. The impact of new technologies such as silicon photomultipliers, gamma imagers and self-shielding directional detectors is presented, and the use of high-performance computing in modelling of system response functions is discussed. Finally, a capability to extrapolate to the location of a source some distance away from a survey trajectory is shown. The extrapolation method includes propagation of the measurement error to the extrapolated region, essential information for nuclear response operators to know if a region is actually clear of radioactivity or not.

1. INTRODUCTION

During a nuclear emergency response operation, one of the most crucial early information products needed is a map showing the locations of any dispersed radioactivity [1] [2]. Typically, national response teams deploy large-volume gamma- and neutron-sensitive detectors, mounted in aircraft, to rapidly produce these broad assessments [3] [4]. Wide-area radiation monitoring is also key to successful on-site operations under the Comprehensive Nuclear-Test-Ban Treaty [5]. Moreover, it is a standard element of a nuclear security operation to perform radiation surveys to establish background levels prior to a major public event [6].

These wide-area search systems are direct descendants of earlier geophysical survey systems used in prospecting for uranium. They are simple, rugged, reliable and offer many advantages. Their high sensitivity permits rapid coverage of very large areas. They are frequently self-contained such that deployment from any available platform is possible. Standardized procedures and data treatment can be applied to deliver an approximate assessment of the dose rate at ground level which can be used to guide ground-based crews.

Despite the advantages of traditional survey systems, they leave a number of capability gaps. No-fly zones can be expected to be encountered in a time-critical nuclear operation, whether a restricted-access site in an on-site inspection, or a zone over which flight is not advised due to the presence of vulnerable assets. A further limitation is that traditional survey systems are direction blind, therefore their measurements can not be extrapolated into an area which was not surveyed. Moreover, time and resource constraints may lead to a decision to gather as much information as possible from a single there-and-back aerial sortie, or from a truckborne survey constrained to roads. In these cases as well, the direction-blind nature of standard mobile survey equipment could lead to a false assessment of a weak source of radioactivity near the survey line when in fact the signal collected was due to a stronger source of radioactivity a distance away.

2. TRADITIONAL CREWED AERIAL SURVEY

In Canada, the Department of Natural Resources (NRCan) is responsible for providing mobile survey capability during a nuclear emergency. In fulfilment of this responsibility, NRCan maintains multiple survey systems, each consisting of large-volume NaI(Tl)-based gamma spectrometers, as well as He-3 neutron detectors. Gamma and neutron data are collected second-by-second and co-registered with geographic information from the Global Navigation Satellite System (GNSS). For more details on the production of radiation map products by NRCan, see [7].

**2.1. Aerial survey of deposited plume after detonation of radiological dispersal device**

NRCan's approach to aerial survey operations in an emergency is well-illustrated by the procedures followed during a Radiological Dispersal Device (RDD) trial which was conducted in Canada in 2012 [8]. Over the course of the trial, three detonations took place each dispersing about 30 to 36 GBq of La-140. The NRCan crew installed a gamma radiation survey system on the exterior of a helicopter, as illustrated in Fig. 1 [9]. The crew then flew a survey of parallel lines at a fixed altitude to map the fallout deposited from the airborne plume following each detonation.

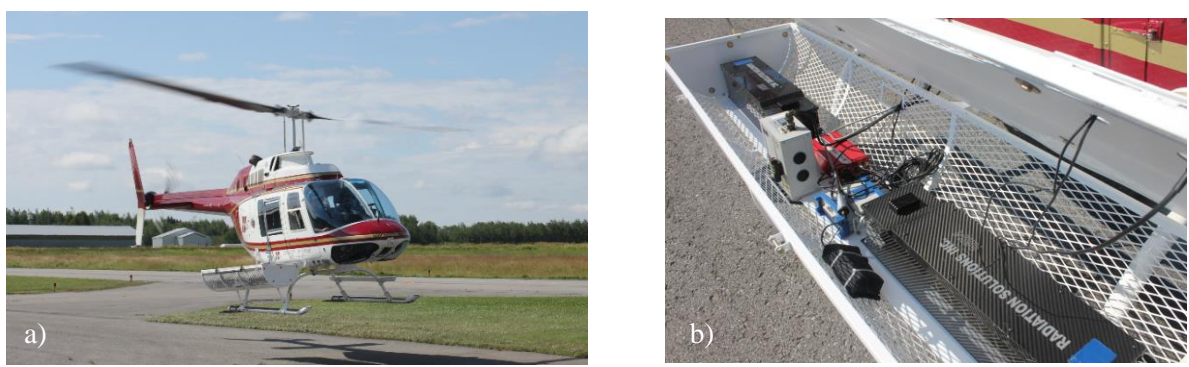

FIG. 1 a) Bell 206 B helicopter survey platform with the radiation measurement system mounted on the exterior in a Dart Aerospace basket. b) Interior of the basket. The black carbon fibre boxes contain two 4 L NaI(Tl) gamma detection crystals. Smaller boxes contain a laser altimeter, an inertial navigation system, and a GNSS receiver. These pictures are reproduced from [9].

In particular, following the third detonation, the NRCan crew flew parallel survey lines spaced 25 m apart at a nominal flying height of 15 m. The resulting map of the fallout was published in [9] and is reproduced here in Fig. 2. It is worthy of note that this contamination map is reported in units of surface area radioisotope concentration. In order to extrapolate from count rate at altitude to radioisotope concentration at the surface of the earth, the effect of an infinite sheet source 15 m below the aircraft had to be understood. Propagation of the radiation through a large volume of air from disc-shaped sources of various radii situated 15 m below the measurement system was simulated in order to infer the system response to an infinite source. Details of this method can be found in [9]. It is a method to calculate the sensitivity of aerial radiometric survey which became possible with the advent of high-performance computing systems consisting of hundreds of central processing units in a parallel configuration.

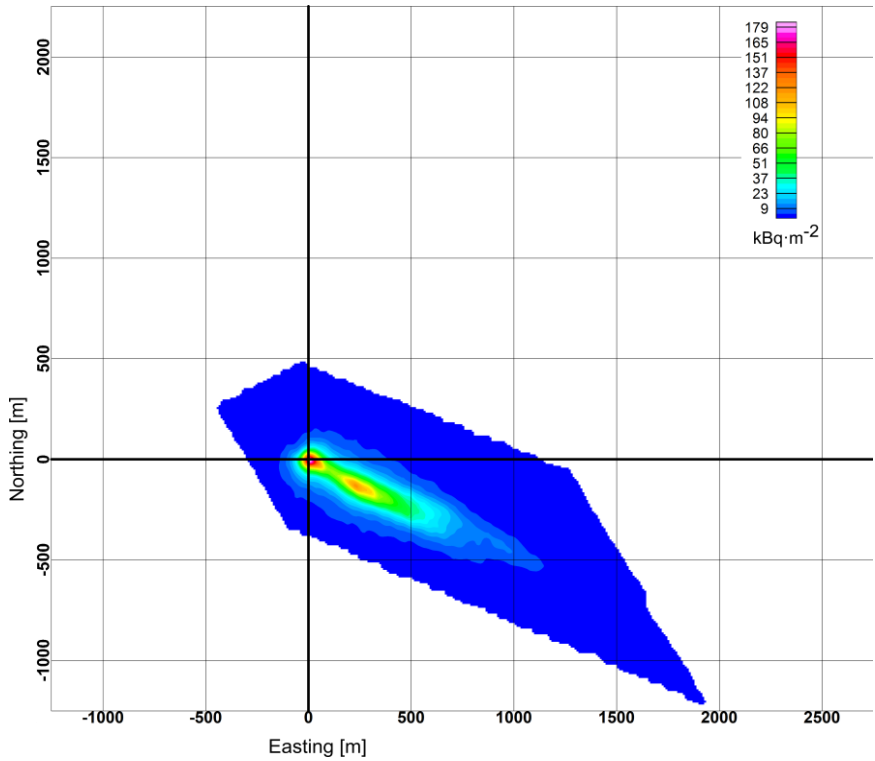

*FIG. 2 Measured distribution of radioactivity deposited following detonation of the third radiological dispersal device. La-140 concentration on the ground is shown averaged over the zone of sensitivity of the aerial system which collected data at 15 m altitude. This figure is reproduced from [9].*

### 2.2. Spatial deconvolution with direction-blind detector

Fig. 3 a) shows a close-up view of the radioactivity distribution in the area around ground zero following the third detonation. A contaminated area of diameter over 100 m is indicated by the aerial survey measurement which was accumulated at 15 m above ground.

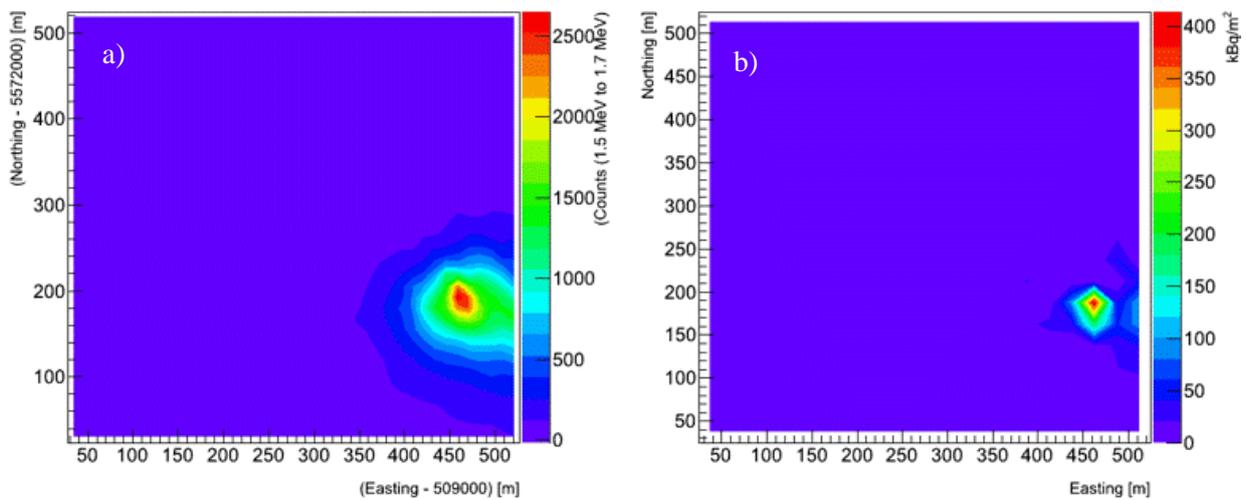

*FIG. 3 a) Close-up of the spatial distribution of the fallout near ground zero after the third RDD detonation, as determined at 15 m altitude. b) The same detailed view of the ground-zero area, after spatial inversion.*

Accumulation of data at altitude causes a smearing effect to the measurement. The aerial measurement system is sensitive to influence from a region on the ground the radius of which grows with the survey altitude [10][11]. Therefore, one must ask how large a contaminated area is indicated on the ground given the area of ~100 m diameter observed at altitude.

NRCan researchers developed a method to deconvolve an aerial survey measurement for the smearing effect of measurement at altitude [12]. The method involves a first step of modelling the transport of radioactivity from distinct regions of space on the ground through the air to the measurement system (these distinct regions of ground will become the pixels of the map-image resulting from the analysis). A second step follows, of performing a minimization to determine which contaminated regions on the ground are necessary and sufficient to account for the distribution of radioactivity measured at altitude. Both of these steps are computationally intensive and have been made possible by advances in high performance computing. The map of the deposited plume from the first RDD blast was deconvolved in this way and published in [12]. The method recovered detail in the plume to about the same level determined by truckborne survey and revealed that the peak contamination at a height of a few metres above the ground was four times that measured at the flying height of 40 m.

Fig. 3 b) shows the distribution of the fallout near ground zero from the third blast, after application of this spatial inversion method. We find that the distribution of the fallout is in fact much narrower, as narrow as 30 – 40 m across, and that the maximum expected concentration is considerably higher, ~ 400 kBq/m$^2$ as compared to 180 kBq/m$^2$ in the undeconvolved measurement. It is important to note that while this underestimation of the strength of hot spots that affects all aerial survey measurements analyzed by traditional methods is significant, it is only relevant for radiation distributions that vary strongly over distance scales smaller or similar to the flying height. Moreover, the underestimation of peak radioactivity does not affect the integrated measurement. Therefore, traditional aerial survey methods are still valuable in providing the gross situational awareness for which they were designed.

## 3. TOMOGRAPHIC RECONSTRUCTION FROM PERIMETER SURVEY USING A SCOTSS GAMMA IMAGER

### 3.1. Silicon photomultiplier-based Compton Telescope for Safety and Security (SCoTSS)

An important method to improve spatial precision in wide-area search has been presented in the previous section, however this method does not provide a way to extrapolate off of a survey line. This problem first became acute for the NRCan nuclear emergency response team when attempting to understand results from truckborne survey constrained to roads. Therefore, the group set about developing high-sensitivity direction-capable and gamma imaging instruments. In this section we present some results obtained using the Silicon photo-multiplier-based Compton Telescope for Safety and Security (SCoTSS) [13][14].

SCoTSS is a Compton gamma imager composed of several hundred individual CsI(Tl) crystals operating in concert to track a gamma ray which scatters within the instrument. By measuring the amounts of energy deposited when the gamma ray undergoes a Compton scatter and then again when it is absorbed, the instrument can reconstruct the angle through which the gamma ray scattered. Thereby a Compton gamma image can be built up, from the measurements of multiple two-hit events. SCoTSS was the first imager of its kind, made possible by advances in solid-state physics which led to the invention of the silicon photomultiplier [15]. Advances in other semi-conductor technologies such as field-programmable gate arrays permitted customization and miniaturization of the electronics, accelerating the transition of this device out of the laboratory and into the field [16].

Fig. 4 a) shows an image of a SCoTSS detector, together with the operator for scale. These are high sensitivity, and therefore large, instruments intended for use as drop-in replacements for traditional mobile survey equipment. Thus, they are capable of mapping radiation spectra second by second while in motion, while providing the additional functionality of giving an image of the radiation emitter. Fig. 4 b) illustrates the high sensitivity of the instrument, and its potential role as a discrete observer in a security operation. In this image an out-patient (not visible) who had been treated with I-131 has been imaged, from several 10's of metres distance, seated in the front seat of their vehicle. Fig. 4 c) is an example of a long-range image. That is an image of a ~35 GBq distributed source of La-140 the centroid of which was several hundred metres from the imager.

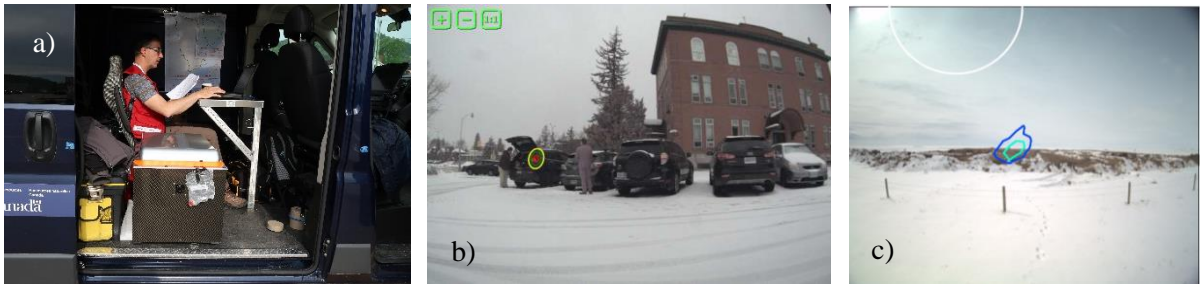

*FIG. 4 a) Image of SCoTSS Compton gamma imager. The imager is the large carbon fibre-clad black box with a silver lid on the floor of the truck between the seated operator and the open door. There is a small plastic bag containing a Cs-137 check source taped to its front face. b) Gamma image taken with SCoTSS of an out-patient (not visible, seated in driver's seat of vehicle) who has undergone treatment with I-131. c) Gamma image taken with SCoTSS of an extended source of ~ 35 GBq of La-140. The La-140 source was extended over an area of approximately 3,200 m$^2$ and the centroid of this distribution was about 300 m from the observer in this image. Image c) is reproduced from [18] with permission.*

### 3.2. Perimeter survey and tomographic reconstruction

A distributed source trial which was conducted in 2018 and described in [17]-[19] serves to illustrate the utility of perimeter survey with a gamma imager in resolving the features of a distribution of radioactivity within a restricted access zone. The gamma image which was presented in Fig. 4 c) is shown again in Fig. 5 a) in geographic coordinates, where the assumption has been made that the radioactive material is situated on the ground.

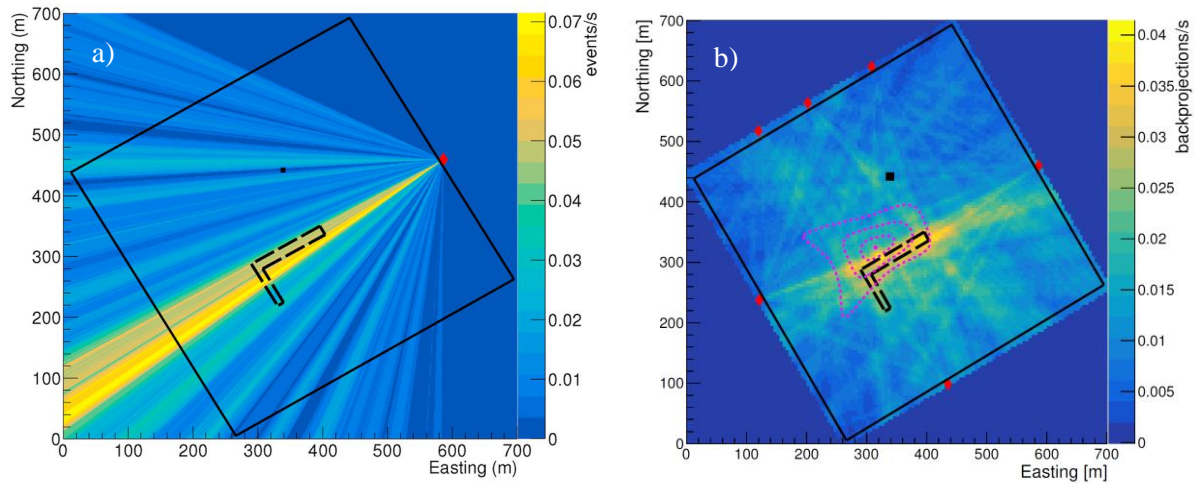

*FIG. 5 Compton images produced by the SCoTSS gamma imager during a dispersed source trial, represented geographically. The dashed black outline shows the approximate L-shaped region over which ~35 GBq of La-140 was sprayed. The solid black square shows the restricted access zone. Red diamonds show the locations from which the SCoTSS imager collected data. The two-dimensional colour distribution shows the rate of tomographic back-projections crossing a pixel. a) Top-down view of a gamma image taken at the position indicated by the red diamond. b) Tomographic image resulting from summing the individual gamma images, one from each of the six view points represented by the red diamonds. Images reproduced from [18] and [19].*

Fig. 5 a) demonstrates the limitation of using information from a gamma image from a single location. The imager gives the two angular components of position (reduced to one by the assumption that the source lies on the ground), but does not specify the third, the distance to the source. Without prior knowledge of the source strength, the source is equally likely to be situated anywhere along the yellow band in Fig. 5 a). To delineate the source, it is necessary to observe it from multiple vantage points. In Fig. 5 b) a tomographic image resulting from the summation of the Compton images from each of the six viewpoints indicated by red diamonds on the restricted access zone perimeter is shown. Fig 5 b) indicates the presence of a long narrow source of radioactivity just offset from the centre of the restricted access zone. In fact, the tomographic image of Fig 5 b) aligns very well with the actual location of the dispersion. The radioactive La-140 was deposited primarily on the long arm of the "L"-shape shown by the dashed black line in Fig. 5. More details of this analysis can be found in [18][19].

It is important to be cognizant of the environmental challenges which affected the gamma imager during this trial. The trial occurred in late winter and the SCoTSS imager and mobile-survey spectrometer was hauled on a tracked, articulated all-terrain vehicle through snowy, muddy, bumpy terrain. The data collection crew were only able to view the source satisfactorily from six locations. Moreover, the signal from the source at the viewing locations for the imager was significantly lower than the signal from the local natural background. The satisfactory performance of the imager in producing a near-real-time product which would provide evidentiary information to inspectors faced with a restricted access site under such severe conditions, is highly promising for its eventual deployment in an on-site inspection under the Comprehensive Nuclear-Test-Ban Treaty, and in other nuclear safety and security operations.

4. ADVANCED RADIATION DETECTOR FOR UNCREWED AERIAL VEHICLE OPERATION (ARDUO)

Remotely piloted aerial systems (RPAS) provide obvious advantages in nuclear emergency response, giving an overhead view of a scene while keeping human operators out of the radiation zone. Researchers at NRCan designed a direction-capable gamma radiation detector for RPAS operations known as the Advanced Radiation Detector for Uncrewed Aerial Vehicle Operations (ARDUO) in order to gather the most information possible given the limited flight endurance offered by contemporary small uncrewed aerial vehicles. ARDUO is shown in Fig. 6 together with Responder, the uncrewed aircraft that carries it. ARDUO is segmented into eight individual CsI(Tl) crystals each independently read out with Silicon Photo-Multiplier (SiPM) light collectors. The gamma-sensitive crystals are arranged in a self-shielding configuration such that direction toward a gamma-emitting source can be calculated once per second and displayed to the operator during flight.

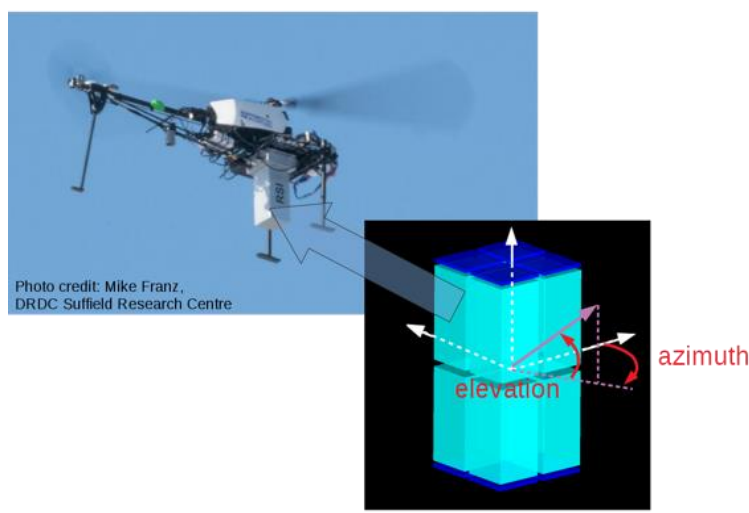

FIG. 6 The Responder main- and tail-rotor remotely-piloted aerial vehicle viewed from below. The ARDUO direction-capable gamma spectrometer is seen mounted below the helicopter body in the box labeled "RSI"[1]. The CsI(Tl) crystals inside the ARDUO are shown by the cyan rectangular prisms in the inset figure. The approximate locations and sizes of the SiPM arrays for light collection are indicated by the small dark blue rectangular prisms on the outside surfaces of the CsI(Tl) crystals. Azimuth angle is measured about the nominal vertical axis of the system, from the direction of the nose of the aircraft. Elevation angle is the angle above (positive) or below (negative) the nominally horizontal plane. This figure reproduced from [20].

---

[1] RSI stands for Radiation Solutions Inc., the equipment manufacturer, located in Mississauga, Canada

### 4.1. Real-time gross count mapping and direction display

Fig. 7 shows the view displayed to the operator during flight of the Responder + ARDUO system. One-second energy spectra provide awareness of the presence and isotopic composition of radioactive sources. This information is displayed in geographic coordinates and updated every second. In addition, the display shows the direction from which radiation is coming with respect to the coordinate system of the aircraft providing an easy signal an intelligent operator (human or artificial) could follow to track or discover a radioactive source in the shortest time possible.

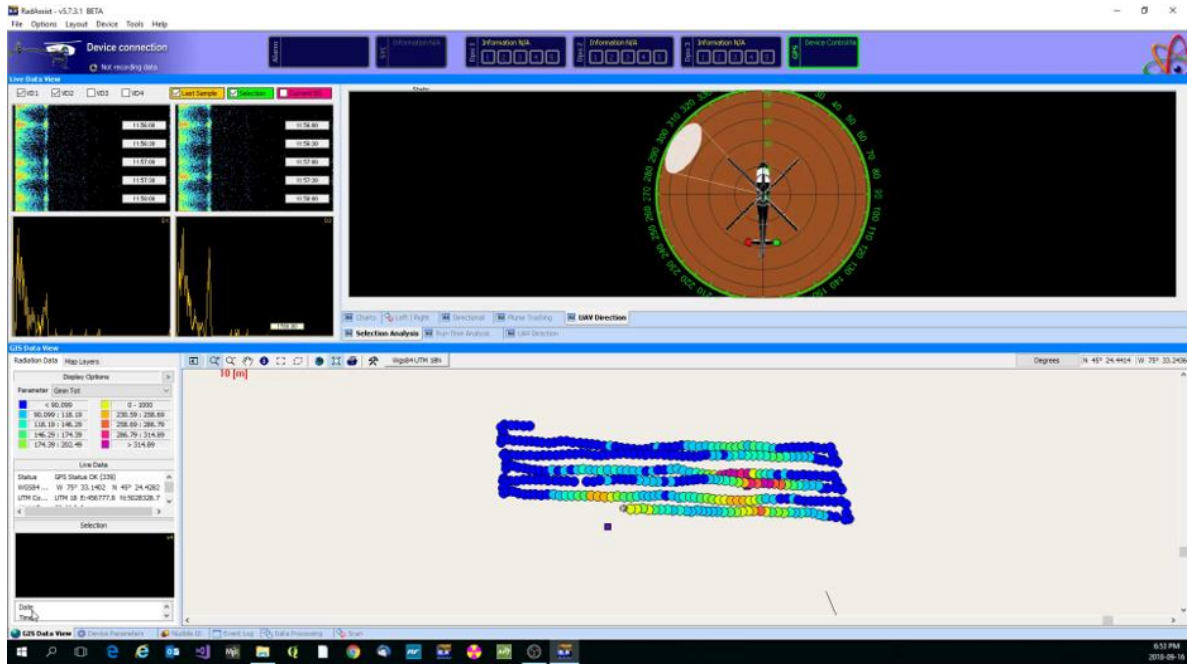

*FIG. 7 Real-time display of the Responder + ARDUO RPAS system. On upper left are one-second spectra for two of the eight crystals, with waterfall plots that show the energy spectra as a function of time. On upper right is shown a top-down view of the result of a direction calculation for a single one-second measurement. In the bottom panel the bread-crumb trail is shown where colour indicates the gross count rate summed over all crystals and each one-second measurement is shown as a function of geographic position. In this flight, two ~160 MBq Cs-137 encapsulated sources had been placed in the terrain at the locations indicated by the purple squares in the map view.*

A particularly interesting scenario for the Responder + ARDUO system is one in which as much information as possible must be gleaned from a single there-and-back sortie. As long as power remains a constraint for aerial survey systems, it is easy to imagine such circumstances, or an equivalent circumstance where there are significant gaps between survey lines in wide-area search. Here, we present a method to extrapolate to a source position away from a survey line. This manuscript contains the first presentation of this method; therefore, we will go into more detail in this and the following sections.

Fig. 8 shows data collected with the Responder + ARDUO system during flights at 10 m height with two ~160 MBq encapsulated Cs-137 sources emplaced in the terrain. Here we neglect all data except for neighbouring flight lines in order to simulate the resource-constrained single-sortie scenario. Considering only the gross count information the data suggest a strong source at Easting ~ 50 m, and perhaps a second weak source at Easting ~ 80 m. Considering the second-by-second direction calculations on the other hand, a different picture is revealed. The direction arrows clearly indicate a source at Easting ~ 80 m where they point downward. These individual one-second direction vectors, however, also indicate a second source somewhere south-eastward of around (Easting, Northing) = (70 m, 30 m). The arrows fail to point correctly to the second source because each direction measurement is being considered individually. What is needed is a method which will determine the optimum placement of sources in the terrain by explaining all of the direction information in the survey, simultaneously.

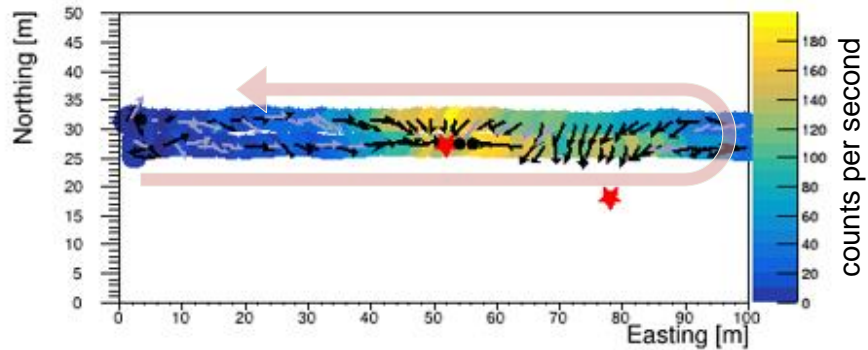

*FIG. 8 Second-by-second measurements taken by the ARDUO RPAS-mounted direction-capable system on one there-and-back sortie. The flight path is schematically indicated by the long pale red curved arrow. The known geographic locations of the two ~160 MBq Cs-137 sources are shown by the red stars. The colour scale shows the count rate determined at each measurement location, plotted geographically. The arrows show the direction calculated at each location, where the tail of the arrow is situated at the measurement location. The arrows include both azimuth and elevation information and are viewed top-down such that longer arrows indicate a direction on the horizon and shorter arrows indicate a direction pointing down or up. Skyward directions are shown in pale blue-grey and downward directions are shown in black. Black dots are arrows pointing directly downward.*

### 4.2. Spatial inversion and extrapolation with direction-capable ARDUO detector

In Fig.9, the black dots display the same data as were shown in Fig. 8, but here the data are shown as a one dimensional histogram, where each bin of the histogram contains the counts accumulated in one second in one geographic location in one of the eight ARDUO crystals.

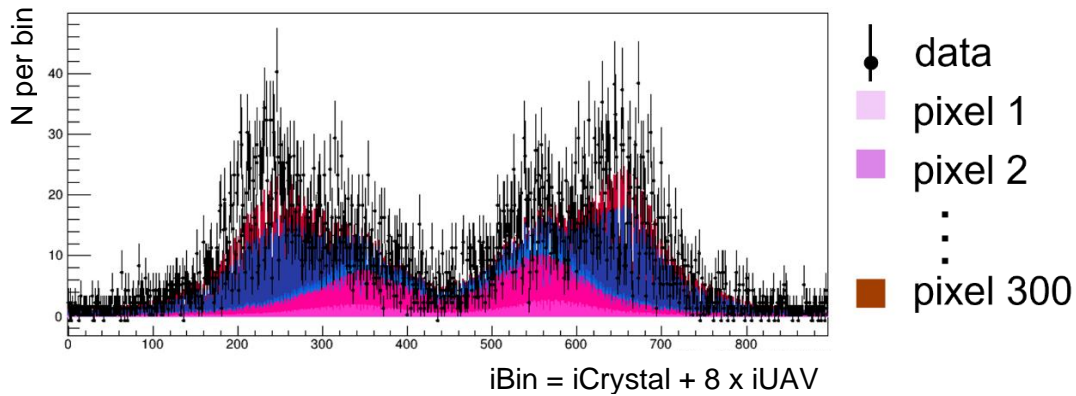

*FIG. 9 Result of fit of the contributions of individual pixels of space to the measured response of the array of crystals. Two ~160 MBq Cs-137 sources are arrayed in the field and the ARDUO RPAS system flew a single out and back sortie. Data are shown by the black dots with statistical uncertainty represented by vertical lines. Coloured histograms show the contributions of up to three hundred different pixels of space. A dominant contribution from two pixels can be seen, shown in dark blue and in magenta. Minor contributions from neighbouring pixels is also indicated in pale blue, pale magenta, and red.*

In a regression method similar to the spatial deconvolution method discussed in Sect. 2, we have first simulated the expected response of an array of ARDUO detectors at the position, altitude, and orientation of the measurement, to individual squares of radioactivity on the ground 4 m × 4 m in area. These square regions of space are the pixels of the map-image that results from the analysis. We then perform a minimization with 300 free parameters, the weights of each region of contaminated space in the 48 m × 100 m terrain, to determine which regions of space are necessary and sufficient to explain the data that was measured. The coloured histograms in Fig. 9 show the result of that fit. The fit demonstrates that most of the measured data can be explained by two pixels, the influences of which are shown by the magenta and dark blue histograms in Fig. 9. The fit suggests that other regions of space contribute at a minor level as well, with the pale magenta, pale blue and red histograms also being necessary to obtain a good representation of the measured data.

The information represented by the coloured histograms in Fig. 9 is shown geographically in Fig. 10 a). Here we see that the fit results in an excellent determination of the location of both the source on the line, and the source a distance from it.

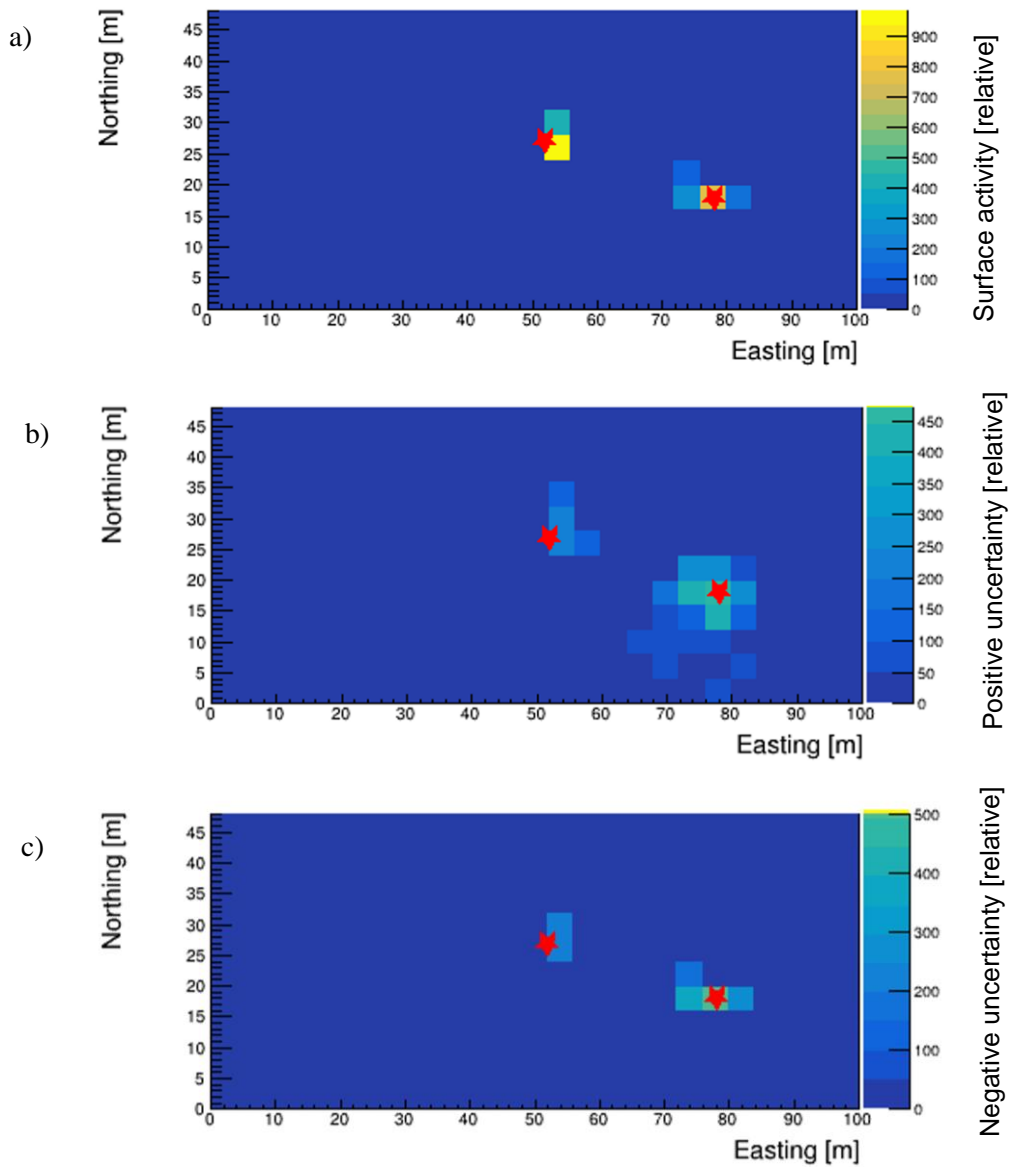

*FIG. 10 Result of spatial inversion of measurements from a single sortie with the ARDUO direction-capable detector. The red stars show the actual locations of the two ~160 MBq Cs-137 sources as determined by geographic positioning system. a) The surface activity concentration resulting from the fit (with an arbitrary scale factor). b) Positive uncertainty on the measured surface activity concentration. c) Negative uncertainty on the measured surface activity concentration.*

Note that the model of the system function is not yet satisfactory enough to use it to determine absolute surface activity concentration. The actual sources were point-like, while the model represents them as distributions over a 4 m × 4 m area. Additionally, dead material enclosing the ARDUO sensitive volume has not yet been included in the model. Therefore, the results in Fig. 10 are shown in relative units. Nevertheless, the advantage of this method in determining source location is clear, especially if one compares the rich information shown in Fig. 10 to the direction indicators which were shown in Fig. 8

Fig.'s 10 b) and c) show the result of propagating the statistical uncertainty of the measurement through the fit. It is clear from Fig. 10 b) that the position of the source which is farther from the line is more uncertain than the position of the source which was under the survey line. The positive uncertainty is also important to consider when asking which regions have been cleared for the ground crews to enter. Fig. 10 b) indicates that there is a spatial region extending as far as (Easting, Northing) ~ (70 m, 5 m) where radioactivity could be present and ground crews should exercise caution. The Fig. 10 c) on the other hand, serves to support the conclusion that radioactivity is present. The Cs-137 signal which was measured is not consistent with a fluctuation from zero radioactivity, according to the negative statistical uncertainty.

5. SUMMARY

It is essential to keep abreast of technological innovations in order to maintain an effective nuclear security architecture. Here, we have shown how several non-nuclear advances have been harnessed in order to improve nuclear and radiation detection and characterization. Advances in high-performance computing have enabled direct calculation of sensitivity functions to convert from the counts collected by an aerial survey system at altitude, to radioactivity concentration on the ground, where previously a laborious series of control flights had been necessary which were subject to flawed assumptions about the similarity of the test strip and survey areas. Semiconductor advances led to the realization of a silicon photomultiplier-based Compton gamma imager, with high-performance computing having also contributed to its design and optimization [21]. The same silicon photomultiplier readout was essential in putting together a compact direction-capable gamma spectrometer with low-voltage power supply for use on uncrewed aircraft. Advanced algorithms are being developed which demonstrate extrapolation to a source location using aerial survey measurements with this detector. With high-performance computing, one can solve for hundreds of free parameters, the contributions of various independent regions of space, simultaneously.

The international community must keep nuclear security high on its agenda. This scientific information about the ways NRCan has made use of emerging technologies and methods in wide-area search for radioactive materials is shared with the intent to strengthen nuclear security globally.

ACKNOWLEDGEMENTS

The authors gratefully acknowledge P.R.B.Saull and A.M.L.MacLeod of the National Research Council of Canada, N.J.Murtha of Alberta Health Services, Canada (formerly of Carleton University, Canada) and A.McCann of Marian College, Dublin, Ireland (formerly of NRCan) for contributions to Compton gamma imaging, particularly the unpublished gamma image Fig. 4 b).